\documentclass[preprint]{aastex}

\begin{document}

\title{Stellar Crowding and the Science Case for Extremely Large Telescopes}

\author{Knut A.G. Olsen\affil{National Optical Astronomy Observatory,
CTIO, Casilla 603, La Serena, Chile\\ kolsen@noao.edu}}

\author{Robert D. Blum\affil{National Optical Astronomy
Observatory, CTIO, Casilla 603, La Serena, Chile\\ rblum@noao.edu}}

\author{Fran\c{c}ois Rigaut} \affil{Gemini Observatory, 670 N. A'ohoku Place,
Hilo, Hawaii, 96720, USA\\ frigaut@gemini.edu}

\begin{abstract}
We present a study of the effect of crowding on stellar photometry.
We develop an analytical model through which we are able to predict
the error in magnitude and color for a given star for any combination of telescope
resolution, stellar luminosity function, background surface brightness, and distance.  We test our predictions with Monte Carlo
simulations of the LMC globular cluster NGC 1835, for resolutions
corresponding to a seeing-limited telescope, the $HST$, and an
AO-corrected 30-m (near diffraction limited) telescope. Our
analytically predicted magnitude errors agree with the simulation
results to within $\sim$20\%. The analytical model also predicts that
errors in color are strongly affected by the correlation of
crowding--induced photometric errors between bands as is seen in the
simulations.  Using additional Monte Carlo simulations and our
analytical crowding model, we investigate the photometric accuracy
which 30-m and 100-m Extremely Large Telescopes (ELTs) will be able to achieve at distances
extending to the Virgo cluster.  We argue that for stellar populations
work, ELTs quickly become crowding-limited, suggesting that
low--Strehl AO systems may be sufficient for this type of science.

\end{abstract}

\section{Introduction}
The study of stellar populations in nearby galaxies has benefited
tremendously from the high sensitivity and resolution of modern
ground-- and space--based telescopes.  Nevertheless, our ability to
study the evolutionary histories of a statistical sample of galaxies
through the properties of their resolved stars is severely hampered by
insufficient resolution to overcome the effects of crowding.  For the
study of the oldest ($\sim$13 Gyr) main sequence stars, even the {\em
Hubble Space Telescope} is limited to the nearest dwarf companions to
the Milky Way, while at the distance of M31 only the brightest stars
or those lying in the lowest surface brightness regions are resolved.

The 30--100m Extremely Large Telescopes (ELTs) now being considered\footnote{e.g.\ the Giant Segmented Mirror Telescope ({\tt http://www.aura-nio.noao.edu/}), the California Extremely Large Telescope ({\tt http://celt.ucolick.org/}), the OWL telescope ({\tt http://www.eso.org/projects/owl/}), the Euro50 telescope ({\tt http://www.astro.lu.se/$\sim$torben/euro50/})}
will provide a giant leap in our ability to study stellar populations
in nearby galaxies.  Using adaptive optics (AO) to correct for the
wavefront distortion produced by the Earth's atmosphere, these
telescopes aim to deliver near diffraction-limited performance at
wavelengths of approximately one to two microns.  Such resolution will vastly
improve our ability to observe faint stars projected against dense
backgrounds.  Nevertheless, even ELTs will be crowding-limited in many
cases.

In this paper, we study the impact of crowding on the stellar
populations science case of ELTs.  The effects of crowding have been
extensively studied by both the optical and radio astronomy
communities, using both analytical and numerical methods.  These
studies have addressed the photometric errors, incompleteness, and
positional errors introduced by crowding (e.g., Scheuer 1957, Condon
1974, Gallart, Aparicio, \& V\'{\i}lchez 1996, Renzini 1998, Hogg 2001,
Stephens et al.\ 2001).  This paper by no means supersedes this
previous work, which as a whole forms an almost complete study of the
subject.  Instead, our goal is to develop a simple set of
tools to help assess the stellar populations science that
may be done with ELTs, focussing particularly on the issue of
crowding-induced photometric errors.  We hope that these tools may
also be useful to a broader set of applications.

In section 2, we develop a simple analytical model of the photometric
effects of crowding.  We test the model against Monte Carlo
simulations (``artificial star tests'') in section 3.  In section 4,
we use further Monte Carlo simulations to demonstrate the performance
of the proposed 30-m Giant Segmented Mirror Telescope (GSMT) in crowded Local
Group environments.  We use our analytical model to determine the
distances out to which 30-m and 100-m AO-corrected telescopes could
perform useful photometry in section 5.  Section 6 contains our
summary and conclusions.

\section{An Analytical Model of Crowding}

For evaluating the possibility of studying stellar populations in
distant galaxies with ELTs, we are particularly interested in
answering the question, ``Given an ELT, to what distance and
background surface brightness can we measure the colors and magnitudes
of stars with $x$\% accuracy?''  We will thus mainly consider the {\em
photometric} effects of crowding.  We will approach the problem by
considering the fluctuations in luminosity produced by a stellar
background, as was done by Tonry \& Schneider (1988) for the case of
distant, unresolved stellar populations.  We note that for the purposes of this study, we ignore all other sources of photometric error, e.g. shot noise, read noise, and flat-fielding errors.  For the environments studied here, crowding is typically the dominant source of error.

A basic limitation of stellar photometry is that there is no way to
remove the star of interest and measure the contribution to its
luminosity of the background underneath.  Instead, the background must
be estimated from adjacent resolution elements, either by averaging
the luminosity within an annulus or by fitting a surface to the
surrounding light.  Thus, the accuracy with which the background can
be estimated (hence the accuracy with which we can measure the light
from the star) is limited by the size of random luminosity
fluctuations within a {\em single} resolution element.  By calculating
the typical amplitude of these fluctuations, we can study
quantitatively the effect of crowding on photometry.

If we are given a luminosity function (LF) $\Phi(l)$ such that $dN = A\Phi(l)dl$, where $dN$
is the number of stars with luminosity between $l$ and $l+dl$ and $A$ is the normalization, then from Poisson statistics the fluctuation in the
luminosity due to shot noise in the number of stars in the interval $l,l+dl$ is
\begin{equation}
\sigma_l = l\sigma_N=l\sqrt{A\Phi(l)dl}
\end{equation}
While performing stellar photometry, either through fixed apertures or through profile fits, we imagine starting with the brightest stars and ``peeling away'' increasingly fainter ones.  Thus, we consider that the background underneath a star of luminosity $L$ in the image receives contributions only from stars with luminosities $l<L$.  
We calculate the typical background fluctuation by summing over the contributions of these fainter stars.  In the limit as the width of the interval $dl$ goes to zero, we find:
\begin{equation}
\sigma_L = \lim_{dl\rightarrow0}\sqrt{\sum\sigma_l^2}=\sqrt{A\int_0^Ll^2\Phi(l)dl}
\end{equation}
Now we need to express $A$ in terms of measurable quantities.  The total background luminosity within the average resolution element is:
\begin{equation}
\Sigma_{\rm res} = A\int_0^\infty l\Phi(l)dl
\end{equation}
If we make the simplifying assumption that only sources within the circular angular resolution element with diameter $a_{\rm res}$ contribute to the light within that element, then the surface luminosity within a 1$^{\prime\prime}$ patch of the sky is:
\begin{equation}
\Sigma_L = \frac{4\Sigma_{\rm res}}{\pi a_{\rm res}^2}
\end{equation} 
This is equivalent to assuming a ``top-hat'' point spread function.
We then express $A$ in terms of the surface luminosity, the angle $a_{\rm res}$, and the first moment of the luminosity function using equations 3 and 4:
\begin{equation}
A=\frac{\pi\Sigma_La_{res}^2}{4\int_0^\infty l\Phi(l)dl} 
\end{equation}
Now we can phrase in mathematical terms the question posed above in words.
We say that, given no other source of errors, confusion--limited photometry of a star with luminosity $L$ requiring fractional accuracy $x$ is constrained by the inequality:
\begin{equation}
\frac{\sigma_L}{L}<x
\end{equation}
Using equation 2 and subsituting the result for $A$ given by equation 5, we find
\begin{equation}
\frac{\sqrt{\pi\Sigma_La_{\rm res}^2\int_0^Ll^2\Phi(l)dl}}{L\sqrt{4\int_0^\infty l\Phi(l)dl}}<x
\end{equation}
Equation 7 is our basic equation describing the effects of crowding on stellar photometry.  We note again that we have assumed a circular ``top-hat'' PSF in our derivation.  A more rigorous approach would be to adopt a realistic PSF and to integrate the contribution of {\em all} stars to any given image coordinate, as was done by e.g. Scheuer (1957).  However, for the PSFs and LFs used here, the integral quickly converges within a small region surrounding the coordinate of interest; within this region, the light added by PSF wings from stars outside the boundary of the resolution element is compensated by the loss of light through the PSF wings from stars within the resolution element,  such that the effect of our assumption of a top-hat PSF is small.
We have also implicitly assumed that stellar clustering occurs only at scales larger than the size of our resolution element.  In the case of star clusters in nearby galaxies, this assumption is valid.  Binaries, on the other hand, violate this assumption; their effect could instead be included in the stellar luminosity function.

Equation 7 will be easier to use if we express the surface luminosity in apparent magnitudes per square arcsec and the stellar luminosity in absolute magnitudes.  Rearranging the equation, we find:
\begin{equation}
\Sigma_m>2M-2.5\log(\frac{4}{\pi}(\frac{\sigma_m}{1.086a_{\rm res}})^2\frac{\int_{M_{lo}}^{M_{hi}}10^{-0.4M^\prime}\Phi(M^\prime)dM^\prime}{\int_{M_{lo}}^M10^{-0.8M^\prime}\Phi(M^\prime)dM^\prime})+(m-M)_\circ
\end{equation}
where $\Sigma_m$ is the surface brightness of the background in magnitudes per square arcsecond, $M$ is the absolute magnitude of the star we want to measure, $M_{\rm lo}$ is the absolute magnitude of the faintest star in the LF, $M_{\rm hi}$ is the absolute magnitude of the brightest star in the LF, $(m-M)_\circ$ is the distance modulus, and $\sigma_m$ is the photometric accuracy in magnitudes.

\subsection{Consequences of the crowding model}
Considering the case of a delta-function LF, $\Phi(M^\prime)dM^\prime = \delta(M^\prime-M)dM^\prime$, equation 8 reduces to:
\begin{equation}
\Sigma_m>M-2.5\log(\frac{4}{\pi}(\frac{\sigma}{1.086a_{\rm res}})^2)+(m-M)_\circ
\end{equation}
which yields the magnitude of the faintest star that can be observed in a region of surface brightness $\Sigma_m$ mags arcsec$^{-2}$ with a given telescope and desired photometric accuracy:
\begin{equation}
m<\Sigma_m-2.5\log(\frac{\pi}{4}(\frac{1.086a_{\rm res}}{\sigma_m})^2)
\end{equation}
If we set $\sigma_m=0.2$, then equation 10 is equivalent to the commonly used rule of thumb that photometry becomes confusion-limited when the background surface brightness equals that produced if the light from the star were spread over 30 resolution elements (e.g. Hogg 2001).  For more accurate photometry, the confusion limit must be taken to be more conservative.

What is the effect of the shape of the luminosity function on the confusion limit?  If we assume a power-law luminosity function $\Phi(l) = l^\alpha$, then for bright stars with $M\sim M_{\rm hi}$, equation 8 becomes:
\begin{equation}
\Sigma_m\gtrsim M-2.5\log(\frac{4}{\pi}(\frac{\sigma_m}{1.086a_{\rm res}})^2)-2.5\log(\frac{3+\alpha}{2+\alpha})+(m-M)_\circ
\end{equation}
for $\alpha>-2$.  Comparing equations 10 and 11, we find that by distributing stars over a range of brightnesses, the surface brightness limit at which one can perform photometry of stars with $M\sim M_{\rm hi}$ with accuracy $\sigma_m$ becomes {\em brighter}.  That is, given a fixed amount of background light, splitting it up into increasing numbers of fainter stars causes the luminosity fluctuations within a resolution element containing a bright star of interest to decrease.  The worst-case scenario is one in which all of the stars have the same brightness as the one we would like to measure, i.e. equation 10.  In the solar neighborhood, $\alpha\sim-1.25$ for $0\lesssim M_V\lesssim 12$ (Binney \& Merrifield 1998), so that the term $-2.5\log(\frac{3+\alpha}{2+\alpha})\sim-0.9$ magnitudes.

To explore the effect of the shape of the luminosity function in greater detail, we use equation 8 to examine the effects of age, metallicity, and the slope of the IMF on crowding.  Figs.\ 1--3 show the surface brightnesses at which photometry is limited by crowding to an accuracy of 10\% in $V$ alongside the $V$ model luminosity functions, which were calculated using Girardi et al.\ (2000) isochrones.  We have assumed a distance modulus of $(m-M)_\circ$=18.5, a resolution of 1\arcsec, and a single-age population for these calculations; the results can be easily translated to other distances simply by adding the difference $(m-M)_\circ-18.5$ to the surface brightnesses, and to other resolutions by adding $-2.5\log(a_{\rm res}^2)$.  Fig.\ 1 shows the crowding-limiting surface brightnesses for stars with $-3<M_V<11$ as a function of age of the environment, assuming solar metallicity, a Salpeter IMF, and a single burst of star formation.  For stars below the main sequence turnoff (which produces a break in the LF at $M_V\sim2-4$), we find that the limiting surface brightness becomes fainter with increasing age.  This is simply the effect of stellar evolution; the number of stars below the turnoff remains unchanged, but the evolution of massive stars causes the environment to dim in brightness.  The spacing of the lines for different values of $M_V$ is set by the local slope of the LF; at ages between 2 and 4 Gyr, the LF is almost flat for $5<M_V<3$, and the lines lie very close to each other.  This is because the photometric error contributed by crowding is dominated by stars with brightness close to that of the star of interest; thus, a flat LF produces roughly equal crowding errors over a large range in magnitude.
Above the main sequence turnoff, the shapes of the curves in Fig.\ 1 are affected strongly by the appearance of the turnoff.  The sharp decrease in the LF means that stars brighter than the turnoff may be measured in brighter environments.  Finally, stars with $M_V\sim-3$ are particularly easy to measure, since the theoretical luminosity functions cut off at $M_V\sim-1$.  The second term in equation 8 becomes increasingly unimportant, and the crowding-limited surface brightness simply becomes proportional to $2M_V$.  This regime is a bit artificial, however, since normally the star of interest is drawn from the same population as the environment in which it is embedded.  

Fig.\ 2 shows the effect of variations in metallicity on the surface brightness at which one can perform 10\% photometry at a distance modulus of 18.5.  In calculating these curves, we have assumed an age of 5 Gyr and a Salpeter IMF.  
Changing the metallicity has very little effect on the crowding-limited surface brightnesses, except for $M_V\sim-3$, near where the theoretical luminosity functions cut off.

Finally, Fig.\ 3 shows the effect of varying the slope $x$ of the IMF on the crowding-limited surface brightness, where the IMF is given by $dN=Am^{-x}dm$ and $m$ is the stellar mass.  For these calculations, we have used an age of 5 Gyr and solar metallicity.  Changing the IMF slope has dramatic effect on the surface brightness limits for stars below the main sequence turnoff.  This is again caused by the fact that the crowding errors are dominated by fluctuations produced by stars with similar brightness to the one of interest.  Carrying this reasoning through with an example, the curve for $M_V=9$ in Fig.\ 3 is roughly defined by keeping the number of stars with $M_V=9$ constant as the IMF slope is varied.  Making the IMF slope steeper thus decreases the brightness of the environment, while making it flatter increases it, as is seen in the figure.  
By contrast, for bright stars, 
we find that steepening the slope of the IMF makes the surface brightness at which we can perform 10\% photometry slightly brighter.  This behavior is predicted by equation 11, where we set $M\sim M_{\rm hi}$.

Figs.\ 1--3 are useful for estimating the surface brightness above which 10\% $V$-band photometry becomes impossible due to crowding.  Figs.\ 4--6 show similar predictions for $K$-band photometry, again assuming a distance modulus of 18.5 and 1\arcsec resolution.

\subsection{The effect of crowding on colors}
So far, we have considered only the effect of crowding on the measurement of stellar magnitudes.  What about the effect on colors?  To calculate this, we have to account for the fact that the crowding-induced errors are correlated between bands.  That is, given two resolution elements $a_1$ and $a_2$ with corresponding bands $m_1$ and $m_2$, many stars contribute light to {\em both} $a_1$ and $a_2$.  
The crowding-induced variance in the color $m_1 - m_2$ is calculated through the covariance matrix:
\begin{equation}
\sigma^2_{m_1-m_2} = \sigma^2_{m_1} + \sigma^2_{m_2} - 2\sigma^2_{m_1m_2}
\end{equation}
where $\sigma^2_{m_1}$ and $\sigma^2_{m_2}$ are the variances for bands one and two and $\sigma^2_{m_1m_2}$ is the covariance.
The variances $\sigma^2_{m_1}$ and $\sigma^2_{m_2}$ are calculated following equations 1 and 2:
\begin{eqnarray}
\sigma^2_{L_1} = A_1\int_0^{L_1}l_1^2\Phi_1(l_1)dl_1 \\
A_1=\frac{\pi\Sigma_{L_1}a_1^2}{4\int_0^{\infty}l\Phi_1(l_1)dl_1} \\
\sigma_{m_1} = 1.086\sigma_{L_1}/L_1 \\
\sigma^2_{L_2} = A_2\int_0^{L_2}l_2^2\Phi_2(l_2)dl_2 \\
A_2=\frac{\pi\Sigma_{L_2}a_2^2}{4\int_0^{\infty}l\Phi_2(l_2)dl_2} \\
\sigma_{m_2} = 1.086\sigma_{L_2}/L_2
\end{eqnarray}
To calculate the covariance, we take the stars occupying resolution
elements $a_1$ and $a_2$ to follow a bivariate Poisson distribution
with means $N_0+N_a=N_1$ and $N_0+N_b=N_2$, where $N_0$ is the number
of stars in common to $a_1$ and $a_2$.  Then (adopting similar notation as for the single star case) $\sigma^2_{l_1l_2} =
A_{12}l_1l_2N_0$.  In the case of circular resolution elements, the
stars found in the smaller resolution element are completely contained
by the larger, and we find:
\begin{eqnarray}
\sigma^2_{L_1L_2} = A_{12}\int_0^{L_1}\int_0^{L_2}l_1l_2\Phi_{12}(l_1,l_2)dl_1dl_2 \\
A_{12}=\frac{\pi\Sigma_{L_1}\min(a_1^2,a_2^2)}{4\int_0^{\infty}\int_0^{\infty}l_1\Phi_{12}(l_1,l_2)dl_1dl_2} \\
\sigma_{m_2m_2} = 1.086\sigma_{L_1L_2}/\sqrt{L_1L_2}
\end{eqnarray}
where we have arbitrarily chosen to consider the crowding-induced
color errors as a function of $m_1$.  Since $\sigma^2_{m_1m_2}$ is
always positive, the error in color in the crowding-limited case will
always be smaller than the simple sum in quadrature of the individual
magnitude errors.  If $\sigma^2_{m_1m_2}$ is large enough, then one
may be able to measure colors {\em more} accurately than individual magnitudes.

\section{The LMC as a Case Study}
Using equations 8 and 12, we can predict the photometric error due to
crowding for any given telescope resolution, background surface
brightness, distance, and luminosity function.  To test these
predictions, we conducted simulations of a field in the LMC which
contains a globular cluster projected against a dense background of
field stars, and compared them to observations of the real LMC
globular cluster NGC 1835.  In the following section, we demonstrate
the effect of crowding in the LMC under natural seeing conditions, at
$HST$ resolution, and finally at the resolution of an AO-corrected
30-m GSMT.

\subsection{The LMC from the ground}
Figure 7 shows a seeing-limited $V$ image of NGC 1835 and its
surrounding field.  This image was taken by A. Walker as part of an
observing run with the CTIO 1.5-m telescope and Tek 2048 CCD, with the
original aim of calibrating the $HST$ WFPC2 photometry of Olsen et
al.\ (1998). The field of view of the camera is 8\farcm3$\times$8\farcm3.  The images
were taken through CTIO copies of the {\it HST} F555W and F814W
filters on the nights of 1995 January 23-26 under photometric
conditions and in good seeing (FWHM$\sim$1\arcsec).

Figure 7 also shows the $V-I,V$ color-magnitude diagram derived from the
seeing-limited images, plotted in 4 radial bins centered on NGC 1835. 
The photometry was
performed with DAOPHOT/ALLSTAR (Stetson 1987), as follows.  After identifying stars down to a signal-to-noise threshold of 3$\sigma$ and performing aperture photometry out to a diameter of 7\farcs2, we derived PSFs using
$\sim$200 bright stars in each image.  We then computed PSF magnitudes with ALLSTAR, used these magnitudes to subtract all stars except for the PSF stars from the original images, and rederived the PSFs with the neighbor stars removed.  We then recomputed the PSF magnitudes with ALLSTAR, measured aperture corrections using the
PSF stars alone, and 
calibrated the photometry using the transformation equations derived by A. Walker from observations of Landolt (1992)
fields and $\omega$ Centaurus.  

It is clear from the CMDs in Figure 7 that the
photometric error and completeness at a fixed magnitude increase
dramatically towards the center of the star cluster, as is to be expected when crowding dominates the photometric errors.
In order to quantify the effects of the crowding, we performed
simulations using artificial $V$ and $I$ images of the cluster and
field.  These images were produced by combining two theoretical
populations, a 14 Gyr old population with [Fe/H]=$-1.5$ representing
the cluster, the other population having the NGC 1835 field star
formation history derived by Olsen (1999) and the chemical enrichment
history assumed therein.  We selected the individual stars described
by these populations using Girardi et al.\ (2000) isochrones (which we
interpolated in age and metallicity), and assuming a Salpeter (1955)
initial mass function.  We sampled stars from the entire mass range available in the isochrones; the lowest mass stars have masses of 0.15 $M_\odot$ and $M_V\sim11.5$.  The number of stars and their assigned
positions match our observed $V$ surface brightness profile, which we parameterized using a King (1966) model having a 1\farcs8 core radius and concentration ($\log(r_t/r_c)$) 1.7 plus a constant background.
We added
the stars to the artifical images using the DAOPHOT routine ADDSTAR
and the PSFs derived from the true images, and assuming an LMC
distance modulus of 18.5.  The full artificial images, of which we
produced one each in $V$ and $I$, contain $>5\times10^6$ stars down to
$V\sim30$ and include stars with the lowest mass available in the Girardi et al.\ (2000) isochrones, 0.15$M_\odot$.  In order to improve the statistics in the highest surface
brightness regions, we simulated an additional 100 images in $V$ and
$I$ of the central 72\arcsec.  These images are slightly shallower,
with $>4\times10^5$ stars down to $V\sim28$, but still contain more
than 95\% of the light of the same region in the deeper image.  We used these images in our analysis of the photometric errors in the central 72\arcsec.

We performed photometry on the artificial images {\em exactly} as we did for
the 1.5-m images.  Figure 8 shows the CMD obtained from one of the
pairs of deep artificial images alongside the artificial $V$ image
itself.  While there are differences in detail between the artifical
and observed images and CMDs, the simulation reproduces the basic
features of the observations: the size and shape of the
cluster, the field star density, and the broad features of the CMD.
We thus conclude that the simulation is adequate for a realistic study
of crowding in this field.

We identified which of the stars in the input list were recovered by
the simulations by comparing the input and output lists by x and y
position.  We found that 99\% of the recoveries were found within 2
pixels ($\sim$0\farcs5, or one-half times the FWHM) of their input position, 
so we adopted this as the radius
beyond which we considered input stars lost during the simulation.  In
cases where multiple matches were found within 2 pixels, we chose the
pair with the smallest absolute difference in magnitude.  We rejected
stars with input magnitudes $>$22.5, as these appeared to produce only
spurious matches. For the same reason, we also rejected stars which were recovered more than two magnitudes brighter than their input values, as these were exclusively matches with faint stars near the detection limit.  We then computed, as a
function of both $V$ magnitude and position in the frame, the
completeness, the median shift in $V$ and $V-I$ experienced by the
stars, and the standard deviation of $V_{\rm out}-V_{\rm in}$ and
$(V-I)_{\rm out}-(V-I)_{\rm in}$ (using Tukey biweights to limit the effects of outliers), which we call 
$\sigma_V$ and $\sigma_{V-I}$.

\subsection{The LMC with $HST$}

Olsen et al.\ (1998; hereafter O98) presented a $V-I,V$ color-magnitude diagram of
NGC 1835 derived from $HST$ WFPC2 observations.  Figure 9 shows the F555W image of NGC 1835 and the accompanying $V-I,V$ CMDs within 4 annuli centered on the cluster.  We used the results
of artificial star simulations conducted in O98 to further test
our crowding model.  O98 added a set of $\sim$56000
artificial stars with 26$\le V\le 16$, $\sim$500 at a time, to copies of the original images of NGC 1835.   The colors and magnitudes of these stars were chosen to mimic the observed CMDs; thus, their distribution resembles that of an LMC field star CMD.  O98 performed
photometry on the artificial images with DoPHOT (Schechter, Mateo, \& Saha 1993),
and recovered the detected artificial stars by searching for matches
between the input and output lists using a 0.6-pixel search radius.
This approach of using the real images sprinkled with a few artificial
stars at a time differs from the one described in section 3.1.
However, both approaches should adequately measure the completeness
and photometric errors due to crowding.

\subsection{The LMC with the 30-m GSMT}
The design of the 30-m GSMT is
currently under study by A.U.R.A.'s New Initiatives Office.  A
science case and ``point design'' have been developed, which are fully
described in Web documents\footnote{\tt
http://www.aura-nio.noao.edu/book/index.html}. 
The point design
features a primary mirror composed of $>$600 hexagonal segments
producing a 30-m filled aperture.  A multi-conjugate adaptive optics
(MCAO) system intends to provide near diffraction-limited resolution,
with Strehl ratio of $\sim$0.5, in the near-infrared ($JHK$) over a
$\sim$2\arcmin ~field of view.

To generate an artificial GSMT image of NGC 1835, we adopted a
hypthetical 4096 x 4096 near-infrared camera with 0\farcs005 pixels
and read noise of 15 $e^-$ per pixel.  We estimated the system
throughput by scaling an 8-m class telescope to 30-m, including all
mirrors, the atmosphere, and an MCAO module.  At $J$, the total system
throughput is 0.31; at $K$, it is 0.40.  Using the list of stars
produced for the seeing-limited simulation (section 3.1), we added the
appropriate stellar flux to individual pixels in two 4096$\times$4096
arrays representing the $J$ and $K$ images.  The faintest of these stars have $J\sim28$ corresponding to the lowest mass stars available in the Girardi et al.\ (2000) isochrones.  We then convolved the
images with simulated MCAO PSFs, the details of which are described at
{\tt http://www.aura-nio.noao.edu/book/index.html}.  
To describe them briefly, the PSFs have diffraction-limited cores with FWHM of 0\farcs009 in $J$ and 0\farcs015 in $K$.  The Strehl ratios are 0.2 in $J$ and 0.6 in $K$.
These PSFs include the effects introduced by the limited number of
actuators in the deformable mirrors, the limited temporal sampling of
the wavefront, spatial aliasing caused by the limited resolution of
the wavefront sensors, and the estimated optical effects of the primary mirror
segments (tilt and segment-to-segment dephasing).  The PSFs are unrealistic in a few ways, however.
First, we assume that anisoplanatism is completely taken out by
the MCAO system, hence the PSFs do not vary with position in the
image.  The power of MCAO lies in its ability to produce a highly constant PSF across the field of view; but if the PSF changes by small amounts in the
central 1\arcmin ~of the field, it will introduce some small extra photometric error.  Second, the PSFs do not vary with time, which in practice may be the limiting factor for the accuracy of the absolute photometric calibration.  That is, if observations of uncrowded fields are needed to calculate aperture corrections for the crowded fields out to radii of several arcseconds, then time variability of the PSF will produce an impact.  
We consider our neglect of these two effects largely irrelevant to this study, the goal of which is mainly to determine the magnitude levels at which {\em crowding} is the dominant source of photometric error.  Third, because of the large size (2048$\times$2048 pixels) of the PSF arrays, we found it intractable to interpolate the full
PSFs to higher spatial sampling.  Thus, the stars in our simulated
images all appear at the centers of the pixels.  Following a suggestion from the referee, we tested the effect of this simplification by using a 200-pixel wide subsection of the GSMT PSFs, which we interpolated to sub-pixel positions before placing stars in the simulated images.  These PSF subsections contain the cores and diffraction rings, as well as most of the halo structure, but neglect $\sim$10\% of the total light.  The photometric errors recovered from these simulations were identical to those with stars placed only at pixel centers.  We thus conclude that the simplification of placing stars only at pixel centers has no effect on the conclusions presented here.

After the PSF convolution, we produced the final simulated images
assuming that the observation contained a set of 100 0.5--second exposures and 100 10--second
exposures.  These images include a sky background of 16.2 mag
arcsec$^{-2}$ in $J$ and 13.7 mag arcsec$^{-2}$ in $K$,
Poisson noise from the sky and astronomical sources, and a
saturation cutoff of 65535 ADU assuming gain of 2 $e^-$/ADU.  We did not simulate the 200 images directly; we instead simulated the average of the two sets of 100 images, which have photon noise and read noise reduced by a factor of 10 compared to the individual images. 
Following the procedures described in section
3.1, we performed photometry on the images with DAOPHOT/ALLSTAR, including derivation of the PSFs but {\em not} including measurement of aperture corrections.
We merged the short- and long-exposure photometry and determined the
completeness and photometric errors from the recovered artificial
stars.  Figure 10 shows the $K$ image alongside the derived $J-K,K$ CMDs in four annuli centered on the cluster.

\subsection {Results}
Using equations 8 and 12, we computed the predicted crowding-induced photometric errors for the NGC 1835 fields of each telescope as a function of magnitude within 4--6 annuli
centered on the cluster.  These predictions require that we know the distance to NGC 1835 and its surrounding field, the age, metallicity, and surface brightness of the stellar population, and the imaging resolution of the telescope.
For the seeing-limited and GSMT cases, we used
the surface brightness profile, distance modulus, and luminosity
function used to generate the input population.  For the $HST$
simulation, we used the observed surface brightness profile (O98, scaled to match the central surface brightness of Mateo 1987), a
distance modulus of 18.5, and the luminosity function derived from the
Girardi et al.\ (2000) isochrone of a 14 Gyr-old population with
[Fe/H]=$-$1.5.  For the seeing-limited and GSMT cases, we used the average FWHM of the profiles of
stars scattered throughout the images to compute $a_{\rm res}$.  For the $HST$ case, we used
two values of $a_{\rm res}$: one corresponding to the diffraction-limited resolution of the telescope (0\farcs06 for F555W and 0\farcs085 for F814W) and another equal to the 2-pixel resolution of the Planetary Camera, $\sim$0\farcs1 in both F555W and F814W.

\subsubsection{Seeing-limited case}
Figure 11 shows the run of $\sigma_V$ and $\sigma_{V-I}$ with $V$ for
the seeing-limited case, compared with our analytical predictions and with the photometric errors reported by DAOPHOT.
While the errors reported by DAOPHOT grossly underestimate the true
errors, our predictions of the dependence of $\sigma_V$ on $V$ agree
within $\sim$20\% with the simulation results over almost the entire
range of surface brightnesses $\Sigma_V$.  The good agreement between the simulated photometric errors and our crowding model clearly demonstrates that crowding is a dominant source of error in the seeing-limited LMC field.  There are some disagreements, however.  First, as a consequence of our assumption of a top-hat PSF and the hard limits of integration in equation 8, 
the appearance of the horizontal branch produces a
discontinuity in the analytical $\sigma_V$ at $V\sim19$, which is not seen in the simulations.  Second, at high surface brightnesses ($\Sigma_V<17.6$), 
the simulation predicts much smaller photometric errors than does the analytical model.  We note, however, that at these surface brightnesses there is a systematic bias towards recovering stars brighter than their input magnitudes.  Clearly, these input stars are only recovered if they sit on top of large background fluctuations that push them above the detection limit, as seen by the poor fit of the input isochrone to the CMDs.   Although this bias causes stars to be recovered with magnitudes quite different from their input values, the {\em dispersion} $\sigma_V$ in the recovered magnitudes actually decreases, because stars that would have been recovered fainter than their input magnitudes simply aren't detected.  In other words, we are able to sample only a narrow tail of the photometric error distribution through the simulations, making our predictions are invalid.  This bias could also explain the effect seen at faint magnitudes for all surface brightnesses, where taken at face value the
simulation results suggest decreasing photometric errors with
increasing magnitude.  The low completeness at these magnitudes insures that only the tail of the photometric error distribution is observed.

Our predictions of the run of $\sigma_{V-I}$ with $V$ are also in
reasonable agreement with the simulation. Both the simulation and our
analytical predictions show that $\sigma_{V-I}$ is {\em lower} than
$\sigma_V$ at fixed $V$, which we attribute to the correlated
crowding-induced photometric errors in $V$ and $I$.  Again, the
DAOPHOT errors tend to underestimate the true errors.  

\subsubsection{HST case}
Figure 12 shows $\sigma_V$, $\sigma_{V-I}$, and the CMDs for six annuli
centered on NGC 1835 as derived from the $HST$ photometry.  As in
 the seeing-limited case, the errors reported by DoPHOT grossly
underestimate the true $\sigma_V$.  Our predicted $\sigma_V$ values,
on the other hand, again generally agree with the simulation results
to within $\sim$20\%.  The two resolutions (diffraction-limited resolution, 0\farcs06 in $V$ and 0\farcs085 in $I$, and 2-pixel, 0\farcs1 PC resolution) used to calculate the
predictions bracket the simulation results, demonstrating that WFPC2's
undersampling of the PSF affects the crowding-induced
photometric errors.  Undersampling likely also explains why the DoPHOT
photometry never achieves an accuracy higher than $\sim$5\% at bright magnitudes, where our crowding model predicts that better performance should
be possible; similar results have been found by others (e.g. Dolphin 2000).

As for the seeing-limited case, our predictions of $\sigma_{V-I}$ for the $HST$ case agree
with the simulations in producing colors that are more accurately
measured than magnitudes.  However, the errors in color predicted by DoPHOT agree better with the simulation results than do our predictions, which include only the effects of crowding.  Thus, for the $HST$ case, we conclude that sources other than crowding dominate the color error budget.

\subsubsection{GSMT case}
Figure 13 shows $\sigma_J$, $\sigma_{J-K}$, and the $J-K,J$ CMDs for
the GSMT LMC cluster simulation.  Adopting $\lambda/D$ as the diffraction-limited resolution, i.e. 0\farcs009 in $J$ and 0\farcs015 in $K$, our analytical predictions suggest that crowding dominates the errors in both magnitude and color.  As for the other simulations, the errors reported by DAOPHOT tend to underestimate the true errors.
 
We note that the high photometric accuracies ($\lesssim$ 1\%) that we achieve with our GSMT photometry for the bright, uncrowded stars are unrealistic.  For these stars, we expect that errors in the PSF model should dominate the error budget.  Esslinger \& Edmunds (1998), based on both simulations and observations with the ADONIS AO system on the ESO 3.6-m telescope, found typical random photometric errors of $\sim$0.05 magnitudes for uncrowded stars.  While MCAO should significantly improve the minimum achievable photometric error, we adopt 5\% minimum photometric errors in the worst-case scenario.  For most applications, such an error will not significantly affect the ability to measure the ages and metallicities of stellar populations.

\subsubsection{Completeness}
Figure 14 demonstrates the intimate connection between
photometric error and completeness, as determined from the simulations
of all three telescopes (seeing-limited, $HST$, and GSMT).  For the seeing-limited case, we have excluded
the regions with $\Sigma_V\le17.6$ where the simulation becomes biased towards lower photometric errors, as discussed above.  We expect that completeness
should depend on the details of the object detection algorithm and the fit
of the PSF to the objects.  Indeed, we see slightly different
shapes to the curves in Figure 14 for the three sets of simulations.  However, Figure 14 demonstrates that in observations of crowded fields with sufficient exposure time to render photon noise negligible, completeness
is dominated by the difficulty of detecting and measuring
objects against the fluctuating background, independent of the shape
of the PSF and the details of the analysis.  From Figure 14, we arrive at a useful rule of thumb: completeness drops sharply when photometric errors due to
crowding exceed $\sim$0.1 magnitudes.

In summary, our simple analytical model of
crowding is able to predict the magnitude error due to crowding in any
environment and observed with any resolution to an accuracy of
$\sim$20\%.  It is also able to reproduce the observation that
colors are measured more accurately than expected, although the
calculation of color error is fairly sensitive to the input
parameters.  Our model demonstrated that our GSMT simulations realistically reproduce the effects of crowding, in spite of the idealizations made.

In the next sections, we compare our crowding model to photometry measured with a real AO system.  We also explore further what ELTs, such as
the GSMT, can do in the study of resolved stellar populations.

\section{M32 with Gemini North + Hokupa'a, NGST, and GSMT}
\subsection{Photometric errors in M32 with Gemini + Hokupa'a}
M32 is a common benchmark for population synthesis studies of more
distant elliptical galaxies, and thus will be an important target for
ELTs.  Integrated spectroscopy combined with population synthesis
analyses conclude that the inner region of M32 is intermediate
($\sim$4$-$5 Gyr) in age and has solar metallicity (Trager et al.\
2000, del Burgo et al.\ 2001).  Davidge et al.\ (2000; hereafter D00) observed the
central 20\arcsec of M32 with Gemini North and Hokupa'a, and
detected AGB stars to $M_K\sim-5$.  While the $\sim$0\farcs12 resolution of their images was insufficient to test whether the population synthesis models are
correct, they established that there are no radial trends in the luminous AGB population within the inner 20\arcsec.

Figure 2 of D00 shows a decrease in completeness and increase in photometric error at fixed magnitude with rising surface brightness, demonstrating that crowding is a dominant source of photometric error in the Hokupa'a M32 photometry.  Moreover, the inner 2\arcsec contains almost purely blended stars, such that crowding prohibits any useful photometry.  

How does the photometric error due to crowding compare with the other sources of photometric error, such as imperfections in the PSF fits?
To answer this question, we performed artificial star tests using D00's M32 images.  First, we followed the prescription for performing photometry outlined by D00 and references therein to demonstrate that we could reproduce the D00 CMDs.  In brief, we removed the variable unresolved background by median-filtering the $H$ and $K$ images through a 100 pixel wide filter and subtracting these from the originals.  We then performed photometry with DAOPHOT/ALLSTAR, as done in Section 3.
However, we did not compute our own aperture corrections; instead, we applied offsets to the $H$ and $K$ magnitudes to bring them to agreement with the D00 photometry, which Tim Davidge kindly provided to us.
Next, we added 14500 artificial stars drawn from a mix of 1 Gyr ([Fe/H] = 0), 5 Gyr ([Fe/H] = 0), and 10 Gyr ([Fe/H] = -0.3) populations to copies of the original images.  The 1, 5, and 10 Gyr old stars were assumed to
comprise 10\%, 45\%, and 45\% of the total by mass, respectively,
roughly following the estimated age distribution of Grillmair et al.\
(1996); we adopted a Salpeter IMF for these populations.  We added the stars to the $H$ and $K$ images 250 at a time, following a spatial distribution mimicking that of M32 itself, and performed photometry on the modified images exactly as done for the originals.  We identified the stars recovered in the simulation by matching the output star list to the input star list, using a 1-pixel search radius.

Figure 15 shows the Gemini+Hokupa'a M32 CMD that we derive within the annulus between 7\farcs4 and 13\farcs1 radius centered on M32.  We have excluded stars with $y$ pixel positions greater than 700, because in this region we found the effects of anisoplanatism on the PSF to be severe and difficult to model.  Figure 15 also shows the photometric errors $\sigma_H$, $\sigma_K$, and $\sigma_{H-K}$ derived from the artificial star tests, along with our analytical predictions.  The predictions assume the same mix of stellar populations used in the artificial star tests, a distance modulus of 24.3 for M32, and 0\farcs12/0\farcs14 resolution in $H$/$K$ (D00).  We find that the photometric errors are consistent with being entirely due to crowding down to a level of $\sim$5\%.  We thus claim that other sources of random photometric error must be $\lesssim$5\%.  This test demonstrates that despite the low Strehl of its PSFs, Gemini+Hokupa'a photometry in crowded fields is limited by the resolution of the PSF core.

\subsection{Simulated NGST performance}
The Next Generation Space Telescope (NGST) aims to provide diffraction-limited performance in the near-infrared with an $\sim$8-m aperture.  Because of the low background in space and the $\sim$100\% Strehl ratio of the PSF, the NGST will provide huge gains in sensitivity over AO-corrected ground-based telescopes of similar size.  However, for crowding-limited photometry, where the limiting factor is the size of the diffraction-limited core of the PSF, the NGST will provide no such gain.  To demonstrate this, we simulated an NGST observation of a field in M32 covering the region observed by D00.   To generate the input star list, we
assumed the same populations used in the Gemini+Hokupa'a artificial star tests, i.e. 10\% 1 Gyr ([Fe/H] = 0), 45\% 5 Gyr ([Fe/H] = 0), and 45\%10 Gyr
([Fe/H] = -0.3) stars.  The faintest stars in the simulation have $K\gtrsim27$, well below the expected completeness limit set by crowding.  We added these stars to 586$\times$586 arrays according to the spatial profile of Kent (1987), assuming 0\farcs035 pixels, which is appropriate for the proposed NIRCam\footnote{{\tt http://www.stsci.edu/ngst/instruments/nircam/}}.  As we did with the GSMT simulations, we added all of the stars to the centers of pixels.  We then convolved these arrays with Krist's (1999) $J$ and $K$ PSFs, which were calculated for an 8-m diameter NGST.  These PSFs have diffraction-limited cores of 0\farcs032 at $J$ and 0\farcs057 at $K$; as such, our pixel scale results in severe undersampling of the $J$ PSF.  We interpolated these PSFs to the 0\farcs035 pixel scale before convolution.

We performed photometry on the simulated images and analyzed the output star list following the procedures outlined in Section 3.  Figure 16 shows the CMD in the 7\farcs4--13\farcs1 annulus centered on M32 compared with the input isochrones.  The CMD is deeper than that of Gemini+Hokupa'a, as is to be expected from the higher resolution of NGST.  
The photometric errors appear consistent with being dominated by crowding, as seen by the comparison of the simulation results with our analytical predictions in Figure 16. 
These predictions adopt as input the properties of the artificial populations; we use the diffraction-limited resolution for $K$ but two resolutions for $J$, 0\farcs035 and 0\farcs070.  The $J$ photometry is clearly affected by the undersampling, as the simulation results fall closer to the 2-pixel resolution curve than the diffraction-limited resolution curve.

\subsection{Simulated GSMT performance}
Because of the effects of crowding, previous studies of M32 leave the nature
of M32's underlying stellar population highly uncertain.  With its much higher resolution, the GSMT can make a large impact on our understanding of this galaxy.  To study the GSMT's performance further, we simulated a GSMT observation of the center of M32.

We generated
simulated $J$ and $K$ images of M32 using the same input star list and spatial distribution as for the NGST simulation.  We added the stars to 4096$\times$4096 arrays, convolved the
arrays with our simulated GSMT PSFs, and scaled the resulting $J$ and
$K$ images to exposure times of 5$\times$10s and 15$\times$120s
using the system efficiencies from Section 3.3.  We analyzed these
images with DAOPHOT, as was done for the LMC GSMT simulation.

Figure 17 shows the simulated $J-K,K$ CMD in the 7\farcs4--13\farcs1 annulus centered on M32, which may be compared to Figs.\ 15 and 16, and the photometric error profiles $\sigma_J$, $\sigma_K$, and $\sigma_{J-K}$.  The depth of the photometry ($J\sim26.5$) is comparable to the faintest magnitude of the input population ($J\sim27.6$).  However, because there is reasonable agreement of the analytical model with the simulation and we understand that the effects of crowding are dominated by fluctuations produced by stars with similar brightnesses to the star of interest, we conclude that the input star list is sufficiently deep to adequately model the effects of crowding.  The simulation shows that near the center of M32, the GSMT will detect stars with $M_J\sim0.5$, below the turnoff of a 1 Gyr population.  Farther from the dense center, such as in the region studied by Grillmair et al. (1996), the GSMT will also detect the older populations.  If a 5 Gyr-old population is present, as suggested by the population
synthesis models, then a telescope such as the GSMT will be able to
observe it.  Compared to the NGST, the GSMT will reach depths $\sim$3 magnitudes deeper.

\section{Beyond the Local Group: 30-m vs.\ 100-m, IR vs. optical}
An important goal for stellar populations research is to measure the
ages, abundances, and kinematics of stars in a statistically
significant sample of galaxies covering a range of morphologies and
environments, so as to overcome the limitations of cosmic variance
(Wyse et al.\ 2000).  How will the performance of a 30-m telescope
compare to that of a 100-m telescope?  The high surface brightnesses
of most galaxies means that crowding will dominate the photometric
errors; equation 8 may thus be used to assess the relative
performances of different apertures and operating wavelengths.

Using equation 8, we calculated crowding limits for 30-m and 100-m
apertures assuming diffraction-limited resolution in $V$ and $K$.
We assumed a stellar surface brightness of $\Sigma_V=22, \Sigma_K=19$
and a 14 Gyr-old population with [Fe/H]$=-1$.  As discussed in Section 2, using other
metallicities and ages could alter the results considerably.
Figures 18 and 19 show the limits at which $\sigma=0.1$ photometry may
be performed with 30-m and 100-m telescopes, respectively, at the
distances of the LMC, M31, the Sculptor group, Cen A, and the Virgo
cluster.  Our limits show that near the half-light radii of giant
elliptical galaxies in the Virgo cluster, the 30-m GSMT will produce
results similar to what is now possible in the core of M32.  CMDs
could be produced down to the base of the giant branch in regions of
the M31, bulge and disk, down to the horizontal branch in the disks of
Sculptor group galaxies such as NGC 253, and to below the tip of the
red giant branch in Cen A.  To reach the old main sequence turnoff in
a galaxy as distant as M31, the GSMT will be restricted to working in
regions of surface brightness $\Sigma_K\gtrsim21$.  If
diffraction-limited resolution were possible at wavelengths as short
as 5000\AA, then the age and metallicity distribution of the M31 bulge
and disk could be measured, the star formation histories of Sculptor
group galaxies would be accessible to study, and the horizontal branch
detected in Cen A.  In the Virgo cluster, we could observe the TRGB to
almost the half-light radii of giant ellipticals.

A 100-m telescope, such as the Overwhelmingly Large telescope (OWL)
being studied by the European Southern Observatory, operating in the
near-infrared will accomplish roughly what a diffraction-limited 30-m
telescope can do in the optical.  With diffraction-limited resolution
in the optical, the 100-m could measure the complete star formation
histories and abundance distributions in galaxies out to the distance
of Cen A; it could even reach the old main sequence turnoff in low
surface brightness regions ($\Sigma_V\sim25$) of Virgo cluster
galaxies.

\section{Summary and Conclusions}
In this paper, we have developed and used an analytical model to
predict the magnitude errors produced by crowding with an accuracy of
$\sim$20\%.  We have tested our predictions with Monte Carlo
simulations of astrophysical environments covering several magnitudes
of surface brightness, as observed with telescopes spanning two orders
of magnitude in resolution.  We have shown that correlations between
the crowding-induced errors in different photometric bands leads to
significantly smaller errors in color than would otherwise be
expected.  Using both Monte Carlo simulations and our analytical
predictions, we have shown to what level 30-m and 100-m AO-corrected
telescopes will perform useful photometry for galaxies at distances as
large as the Virgo cluster.

Throughout, we have neglected the amount of integration time needed to
reach the desired magnitude.  How long will it take for a given ELT to
reach its crowding limit?  From equation 8, we see that the magnitude
limit of a telescope for crowding-limited stellar photometry is
proportional to the area of a resolution element.  For a
seeing-limited telescope, we reach the obvious conclusion that
building a bigger telescope does not improve the crowding limit;
however, the time needed to reach the crowding limit decreases
proportionally to the area of the aperture.  On the other hand, if the
telescope is diffraction-limited, as is expected to be the case for
AO-corrected ELTs, the depth of the crowding limit scales roughly with
the collecting area of the telescope, neglecting the effect of the
detailed shape of the luminosity function.  Since the flux received
also scales with the collecting area, this means that the time it
takes for a diffraction-limited telescope to reach its confusion limit
is independent of aperture size.  As an example, from section 3.4 we
found that WFPC2 with $HST$ is limited to $\sigma_V=0.1$ at
$V\sim23.5$ when the stellar background contributes 21.0 mag
arcsec$^{-2}$, which is the typical surface brightness in spiral arms
(e.g. Okamura 1988) and at the effective radius of giant elliptical
galaxies (e.g. Caon et al.\ 1990).  According to the WFPC2 exposure
time calculator, the exposure time needed to achieve S/N=10 at V=23.5
against a background of 21.0 mag arcsec$^{-2}$ is $\sim$90 seconds.
Thus, a diffraction-limited telescope is {\em rapidly} limited by
crowding in regions with surface brightnesses typical of normal
galaxies.  Even in the infrared observed from the ground, a GSMT
with Strehl of 1.0 would reach its crowding limit in $<300$ seconds when 
observing a field
with $\Sigma_K=18$ mag arcsec$^{-2}$~\footnote{calculated using IRAF CCDTIME, 30-m primary with 1.5-m central obscuration, and MCAO imager as described in text}.  The rapid approach of the
crowding limit argues that the Strehl ratios of ELT PSFs need not be
large for crowded-field photometry.  The huge difference in spatial
scale between the diffraction-limited core and the seeing-limited
``halo'' for ELTs means that lower Strehl simply raises the sky background and
increases the exposure time needed to reach the crowding limit.  As
seen here, diffraction-limited telescopes operating in crowded
environments have plenty of exposure time to spare.

On a more general note, artificial star tests are useful and necessary for many areas of stellar populations research, e.g. measuring LFs in clusters and determining the star formation histories of galaxies from field CMDs.  However, as demonstrated here and elsewhere, care must still be taken to ensure that the simulations faithfully represent reality.  The tools developed here may provide a sanity check on the photometric errors and completeness levels determined from simulations.  For some applications, one may even be able to replace the artificial star tests with our analytical crowding estimates.

\acknowledgements
This work grew out of discussions held at the NOAO GSMT stellar populations panel, which was chaired by Rosie Wyse.  We are grateful for her leadership of this panel and for follow-up discussions with her afterwards.  We also thank Jay Frogel, Mario Mateo, Joan Najita, Andrew Stephens, and Steve Strom for their input and wisdom.  We thank Tim Davidge for kindly providing us with his M32 photometry.  We are indebted to the anonymous referee, who provided a thorough and thought-provoking report which substantially improved this paper.  Finally, KO thanks Dara Norman for continued inspiration.

\newpage
\begin{figure}
\caption{Left: $V$-band luminosity functions for solar metallicity stellar populations from the Girardi et al. (2000) models, assuming a Salpeter IMF.  Each luminosity function is labelled with the assumed age, which runs from 1 to 10 Gyr.  The curves have been multiplied by arbitrary scale factors for display purposes.  Right: The surface brightnesses at which crowding limits photometry to 10\% accuracy, displayed as a function of age for stars with $11 \le M_V \le -3$.  The calculation assumes spatial resolution of 1\arcsec ~and a distance modulus of 18.5.  To translate the predictions to other resolutions $a_{\rm res}$ or distances $(m-M)_0$, one simply needs to add constants of $2.5\log(a_{\rm res}^2)$ and $(m-M)_0-18.5$ to the values on the $y$ axis.}
\end{figure}
\begin{figure}
\caption{Left: $V$-band luminosity functions for 5 Gyr stellar populations from the Girardi et al. (2000) models, assuming a Salpeter IMF.  Each luminosity function is labelled with the assumed metallicity.  The curves have been multiplied by arbitrary scale factors for display purposes.  Right: The surface brightnesses at which crowding limits photometry to 10\% accuracy, displayed as a function of metallicity for stars with $11 \le M_V \le -3$.  As in Fig.\ 1, the calculation assumes spatial resolution of 1\arcsec ~and a distance modulus of 18.5.}
\end{figure}
\begin{figure}
\caption{Left: $V$-band luminosity functions for 5 Gyr solar metallicity stellar populations from the Girardi et al. (2000) models.  Each luminosity function is labelled with the assumed slope of the IMF, where 2.35 corresponds to a Salpeter IMF.  The curves have been multiplied by arbitrary scale factors for display purposes.  Right: The surface brightnesses at which crowding limits photometry to 10\% accuracy, displayed as a function of IMF slope for stars with $11 \le M_V \le -3$.  As in Fig.\ 1, the calculation assumes spatial resolution of 1\arcsec ~and a distance modulus of 18.5.}
\end{figure}
\clearpage
\begin{figure}
\caption{Left: $K$-band luminosity functions for solar metallicity stellar populations from the Girardi et al. (2000) models, assuming a Salpeter IMF.  Right: The surface brightnesses at which crowding limits photometry to 10\% accuracy, displayed as a function of age for stars with $7 \le M_K \le -5$.  See Fig. 1\ caption for details.}
\end{figure}
\begin{figure}
\caption{Left: $K$-band luminosity functions for 5 Gyr stellar populations from the Girardi et al. (2000) models, assuming a Salpeter IMF.  Right: The surface brightnesses at which crowding limits photometry to 10\% accuracy, displayed as a function of metallicity for stars with $7 \le M_K \le -5$.  See Fig. 1\ caption for details.}
\end{figure}
\begin{figure}
\caption{Left: $K$-band luminosity functions for 5 Gyr solar metallicity stellar populations from the Girardi et al. (2000) models.  Right: The surface brightnesses at which crowding limits photometry to 10\% accuracy, displayed as a function of IMF slope for stars with $7 \le M_K \le -5$.  See Fig. 1\ caption for details.}
\end{figure}
\begin{figure}
\caption{NGC 1835 from the ground.  {\it Left:} This $V$ image was taken by A. Walker in January 1995 with the CTIO 1.5-m telescope and Tek 2048$\times$2048 camera.  The resolution is $\sim$1\arcsec and the field of view is $\sim$8\arcmin$\times$8\arcmin. {\it Right:} $V-I,V$ color-magnitude diagram from the CTIO 1.5-m image displayed on the left, shown in four annuli centered on NGC 1835; the limits of the radial bins are labeled in the plots.  The inner annuli contain predominantly cluster stars, while the outer annuli are dominated by the LMC field.}
\end{figure}
\begin{figure}
\caption{Left: a simulated seeing-limited image of NGC 1835.  Right: $V-I,V$ CMD derived from the simulated image shown on the left, displayed in the same four annuli centered on the cluster shown in Fig.\ 7.}
\end{figure}
\clearpage
\begin{figure}
\caption{Left: HST WFPC2 F555W image of NGC 1835, from O98.  The resolution is $\sim$0\farcs1 and the field of view is $\sim$2\farcm5$\times$2\farcm5.  Right: $V-I,V$ CMD derived by O98 from the HST image, displayed in four annuli centered on NGC 1835, as labeled in the plots.}
\end{figure}
\begin{figure}
\caption{Left: Simulated $K$ image of NGC 1835 with the 30-m Giant Segmented Mirror Telescope.  The resolution is $\sim$0\farcs015 and the field of view is 20\arcsec$\times$20\arcsec.  Right: $J-K,J$ CMD derived from the simulated image, displayed in four annuli centered on the cluster. }
\end{figure}
\begin{figure}
\caption{Crowding in NGC 1835 from the ground.  The middle column of plots shows the CMDs derived at various annuli in one of the simulated seeing-limited images of NGC 1835, compared with the input isochrone.  The local surface brightnesses are labeled.  In the left-hand column, the points indicate the photometric color errors $\sigma_{V-I}$ as derived from the simulation.  The thick solid lines show our theoretical predictions of $\sigma_{V-I}$, the thin lines the errors reported by DAOPHOT.  In the right-hand column, $\sigma_V$ is plotted instead.}
\end{figure}
\addtocounter{figure}{-1}
\begin{figure}
\caption{cont.}
\end{figure}
\begin{figure}
\caption{Crowding in NGC 1835 with WFPC2 onboard $HST$.  As in Fig.\ 11, the middle column of plots shows the CMDs derived from the observed images in various annuli centered on NGC 1835, compared with a representative 14 Gyr, [Fe/H]=$-1.5$ Girardi et al. (2000) isochrone.  The local surface brightnesses are labeled.  In the left-hand column, the points indicate the photometric color errors $\sigma_{V-I}$ as derived from the simulation.  The thick solid lines show our theoretical predictions of $\sigma_{V-I}$, the thin lines the errors reported by DAOPHOT.  Predictions for both diffraction-limited (red lines) and 2-pixel resolutions (black lines) are shown; these predictions bracket the simulation results, demonstrating the effect of undersampling.  In the right-hand column, $\sigma_V$ is plotted instead.}
\end{figure}
\addtocounter{figure}{-1}
\begin{figure}
\caption{cont.}
\end{figure}
\clearpage
\begin{figure}
\caption{Crowding in the simulated GSMT image of NGC 1835.  As in Fig.\ 11, the middle column of plots shows the CMDs derived in various annuli centered on the cluster, compared with the input isochrone (red).  The local surface brightnesses are labeled.  In the left-hand column, the points indicate the photometric color errors $\sigma_{J-K}$ as derived from the simulation.  The solid black lines show our theoretical predictions of $\sigma_{J-K}$ assuming diffraction-limited resolution (0\farcs009 in $J$, 0\farcs015 in $K$).  The thin black lines show the errors reported by DAOPHOT.  In the right-hand column, $\sigma_J$ is plotted instead.}
\end{figure}
\addtocounter{figure}{-1}
\begin{figure}
\caption{cont.}
\end{figure}
\begin{figure}
\caption{Completeness in our NGC 1835 simulations is plotted against $\sigma_V$ (left) and $\sigma_{V-I}$ (right).  Solid points show values from the seeing-limited simulation, open circles those from the $HST$ simulation, and open triangles those from the GSMT simulation.}
\end{figure}
\begin{figure}
\caption{Gemini+Hokupa'a measurements of M32.  At top left, the CMD derived for the annulus $7\farcs4 \le r \le 13\farcs1$ centered on M32 in the Davidge et al.\ (2000) images is shown.  The remaining three panels show the photometric errors $\sigma_{H-K}$, $\sigma_H$, and $\sigma_K$ derived from artificial star tests (filled circles) compared with our theoretical predictions.  The predictions (thick lines) are for resolutions of 0\farcs12 in $H$ and 0\farcs14 in $K$, while the thin lines show the DAOPHOT-reported errors.}
\end{figure}
\clearpage
\begin{figure}
\caption{Simulated M32 measurements with NGST.  At top left the CMD derived for the annulus $7\farcs4 \le r \le 13\farcs1$ centered on M32 is shown compared with the input isochrones (red lines in electronic version, gray lines in print version).  The remaining three panels show the photometric errors $\sigma_{J-K}$, $\sigma_J$, and $\sigma_K$ derived from the simulations (filled circles) compared with our theoretical predictions.  The black lines show the predictions for diffraction-limited resolution in $J$ and $K$ (0\farcs032 and 0\farcs057 respectively); the red lines (electronic version; dark gray lines in print version) show the predictions for 2-pixel (0\farcs7) resolution in $J$.  The green lines (electronic version; light gray lines in print version) show the DAOPHOT errors.}
\end{figure}
\begin{figure}
\caption{Simulated M32 measurements with GSMT.  At top left the CMD derived for the annulus $7\farcs4 \le r \le 13\farcs1$ centered on M32 is shown compared with the input isochrones (red lines in electronic verision, gray lines in print version).  The remaining three panels show the photometric errors $\sigma_{J-K}$, $\sigma_J$, and $\sigma_K$ derived from the simulations (filled circles) compared with our theoretical predictions.  The black lines show the predictions for diffraction-limited resolution in $J$ and $K$ (0\farcs009 and 0\farcs015 respectively); the green lines (electronic version; light gray lines in print version) show the DAOPHOT errors.}
\end{figure}
\begin{figure}
\caption{Crowding limits with a 30-m AO-corrected telescope.  The lines indicate the magnitudes at which $\sigma\lesssim0.1$ photometry is possible in regions of surface brightness $\Sigma_V=22, \Sigma_K=19$ for galaxies at the indicated distances.}
\end{figure}
\begin{figure}
\caption{Crowding limits with a 100-m AO-corrected telescope.  See Figure 18 for explanation.}
\end{figure}


\newpage
\begin{thebibliography}{}
\bibitem[]{} Binney, J., \& Merrifield, M., 1998, {\it Galactic Astronomy} (Princeton: Princeton University Press), p. 124
\bibitem[]{} Caon, N., Capaccioli, M., Rampazzo, R. 1990, A\&AS, 86, 429
\bibitem[]{} Condon, J. J. 1974, \apj, 188, 279
\bibitem[]{} Davidge, T., Rigaut, F., Chun, M., Brandner, W., Potter, D., Northcott, M., Graves, J. E. 2000, \apj, 545, 89
\bibitem[]{} del Burgo, C., Peletier, R. F., Vazdekis, A., Arribas, S., Mediavilla, E. 2001, \mnras, 321, 227
\bibitem[]{} Dolphin, A. E. 2000, \pasp, 112, 1383
\bibitem[]{} Esslinger, O., \& Edmunds, M. G. 1998, A\&AS, 129, 617
\bibitem[]{} Hogg, D. W. 2001, \aj, 121, 1207
\bibitem[]{} Gallart, C, Aparicio, A.,\& V\'{\i}lchez, J. M. 1996, \aj, 112, 1928
\bibitem[]{} Girardi, L., Bressan, A., Bertelli, G., \& Chiosi, C. 2000, A\&AS, 141, 371
\bibitem[]{} Grillmair, C. J., Lauer, T. R., Worthey, G., Faber, S. M., Freedman, W. L., Madore, B. F., Ajhar, E. A., Baum, W. A., Holtzman, J. A., Lynds, C. R., O'Neil, E. J., Jr., Stetson, P. B. 1996, \aj, 122, 1975
\bibitem[]{} Kent, S. M. 1987, AJ, 94, 306
\bibitem[]{} King, I. R. 1966, AJ, 71, 64
\bibitem[]{} Landolt, A. U. 1992, AJ, 104, 340
\bibitem[]{} Mateo, M. 1987, \apj, 323, L41
\bibitem[]{} Okamura, S. 1988, \pasp, 100, 524
\bibitem[]{} Olsen, K. A. G., Hodge, P. W., Mateo, M., Olszewski, E. W., Schommer, R. A., Suntzeff, N. B., \& Walker, A. R. 1998, \mnras, 300, 665
\bibitem[]{} Olsen, K. A. G. 1999, \aj, 117, 2244
\bibitem[]{} Renzini, A. 1998, \aj, 115, 2459
\bibitem[]{} Salpeter, E. E. 1955, \apj, 121, 161
\bibitem[]{} Schechter, P. L., Mateo, M., \& Saha, A. 1993, \pasp, 105, 1342
\bibitem[]{} Scheuer, P. A. G. 1957, Proc. Cambridge Philos. Soc., 53, 764
\bibitem[]{} Stephens, A. W., Frogel, J. A., Freedman, W., Gallart, C., Jablonka, P., Ortolani, S., Renzini, A., Rich, R. M., Davies, R. 2001, \aj, 121, 2584
\bibitem[]{} Stetson , P. B. 1987, \pasp, 93, 1439
\bibitem[]{} Strom, S. et al. 2002, ``Enabling a Giant Segmented Mirror Telescope for the Astronomical Community,'' {\tt http://www.aura-nio.noao.edu/book/index.html}
\bibitem[]{} Tonry, J. \& Schneider, D. P. 1988, \aj, 96, 807
\bibitem[]{} Trager, S. C., Faber, S. M., Worthey, G., González, J. J. 2000, \aj, 119, 1645
\bibitem[]{} Wyse, R. F. G.  et al. 2000, ``Report of the Stellar Populations Panel,'' {\tt http://www.aura-nio.noao.edu/book/ch2/2\_C.pdf}
\end{thebibliography}
\end{document}